\documentclass{Interspeech}

\interspeechcameraready

\title{Unfolding A Few Structures for The Many: Memory-Efficient Compression of Conformer and Speech Foundation Models }

\author[affiliation={1}]{Zhaoqing}{Li}
\author[affiliation={1}]{Haoning}{Xu}
\author[affiliation={2}]{Xurong}{Xie}
\author[affiliation={3}]{Zengrui}{Jin}
\author[affiliation={1}]{Tianzi}{Wang}
\author[affiliation={1}]{Xunying}{Liu}

\affiliation{}{The Chinese University of Hong Kong}{China}
\affiliation{Institute of Software}{Chinese Academy of Sciences}{China}
\affiliation{Department of Electronic Engineering}{Tsinghua University}{China}
\email{zqli@se.cuhk.edu.hk, xyliu@se.cuhk.edu.hk}

\usepackage{comment}

\usepackage{multirow}
\usepackage{colortbl}  %彩色表格需要加载的宏包
\usepackage{xcolor}
\usepackage{array}
\usepackage{float}
\usepackage{cite}
\usepackage{amsmath}

\usepackage{bbding}
\usepackage{pifont}

\usepackage{IEEEtrantools}

\begin{document}
\bstctlcite{IEEEexample:BSTcontrol}

\maketitle

% the abstract here must exactly match the abstract entered into the paper submission system
\begin{abstract}

This paper presents a novel memory-efficient model compression approach for Conformer ASR and speech foundation systems. Our approach features a unique ``small-to-large" design. A compact ``seed" model containing a few Conformer or Transformer blocks is trained and unfolded many times to emulate the performance of larger uncompressed models with different logical depths. The seed model and many unfolded paths are jointly trained within a single unfolding cycle. The KL-divergence between the largest unfolded and smallest seed models is used in a self-distillation process to minimize their performance disparity. Experimental results show that our foldable model produces ASR performance comparable to individually constructed Conformer and wav2vec2/HuBERT speech foundation models under various depth configurations, while requiring only minimal memory and storage. Conformer and wav2vec2 models with a reduction of 35\% and 30\% parameters are obtained without loss of performance, respectively.

%%   We present a novel architecture compression approach based on a small-to-large perspective, which emphasizes not only model performance but also memory efficiency to address the memory challenges raised by increasingly large automatic speech recognition (ASR) systems. We propose directly optimizing one small model and scaling it up to arbitrary logical depths by flexibly reusing its network structures, while not expanding the physical memory footprint, adapting to the varying requirements of edge devices. We introduce a training framework that enables joint training of models sharing physical layers and integrates a self-distillation process to further improve performance. Experimental results demonstrate that our foldable model achieves performance comparable to individually trained models across multiple depth configurations for both Conformer and wav2vec2 speech foundation models while can significantly reduce memory usage for training and storage.
\end{abstract}

% 1. scalability with number of systems being compressed

% 2. system compression time (training)

% 3. memory requirements during system compression

% 4. performance-wise lossless compression

\section{Introduction}

The field of automatic speech recognition (ASR) systems is witnessing a significant surge in model size, driven in part by the recent proliferation of self-supervised learning (SSL) models such as wav2vec2.0~\cite{baevski2020wav2vec}, HuBERT~\cite{hsu2021hubert}, data2vec~\cite{baevski2022data2vec}, and WavLM~\cite{chen2022wavlm}. This increase creates an urgent demand for effective and efficient model compression techniques. Current compression methods can be broadly categorized into two main approaches: low-bit quantization~\cite{quant1,quant2,quant3,xu2025effective} and model architecture compression. Low-bit quantization reduces the bit width of model weights to reduce storage and computational costs, whereas architecture compression directly decreases the number of parameters in the models.

This paper focuses on architecture compression, which presents several critical challenges that need to be addressed: \textbf{1)} Ensuring model accuracy and effectiveness are maintained after compression. \textbf{2)} Developing a means that can scale the compression process to efficiently generate diverse systems to meet the varying demands of different edge devices. \textbf{3)} Tackling memory consumption throughout the model compression process, as this is vital for the practical application of these techniques in resource-constrained environments. 

To navigate these challenges, a variety of architecture compression techniques have emerged, including knowledge distillation~\cite{rathod2022multi,distilhubert,dist1,dist2,dist3,de2023distilling,park2023conformer}, pruning~\cite{pru3,pru4,pru5,dphubert,jiang2023accurate,lodagala2023pada,hj,wang2023task,gu2024sparsewav}, low-rank matrix factorization~\cite{lr1,lr2,lr3,lilossless}, and weight sharing~\cite{lan2019albert,gao2021extremely,lin2023weight,hernandez2023sharing,wang2024residualtransformer}. Although many of these methods emphasize preserving the performance of compressed models, they typically generate a single compressed model at a time. This requires multiple compression processes and the maintenance of various model sizes to satisfy the unique needs of edge devices, ultimately inflating development and upkeep costs. Some studies have introduced frameworks for jointly training multiple nested weight-sharing systems, which help alleviate training time and reduce storage requirements to a degree~\cite{once,uslim,zhang2021self,nagaraja2021collaborative,wang2022lighthubert,li2024one,akhtar2023small}. However, so far, most methods frequently overlook critical memory considerations during the compression process. For instance, knowledge distillation typically involves a large teacher model, and joint training of multiple systems can even lead to prohibitive memory demands. Consequently, the substantial memory footprint of these approaches can hinder their practical use, particularly when working with large models in environments with limited resources.

To address these challenges, this paper proposes for the first time a novel \textit{small-to-large} perspective on model architecture compression. Specifically, we present a memory-efficient network structure unfolding approach that not only prioritizes model performance and scalability of compression, but also emphasizes memory efficiency in two key aspects: \textit{During training}, we shift the focus away from requiring a large model and instead concentrate on optimizing a small ``seed" model. The model is enhanced by iteratively reusing its network structures to scale up model complexity without expanding the physical memory footprint. \textit{For inference}, a small seed model can be unfolded to arbitrary logical depths, providing the flexibility to adapt to the varying demands of different environments while maintaining minimal storage and memory requirements.

To do this, we propose an effective training framework that supports the joint training of models sharing physical layers and unfolding to different logical depths. In addition, we incorporate a self-distillation process into the training cycle to further enhance performance. This model-agnostic framework is easy to implement and yields versatile models that are well-suited for real-world applications. Experimental results validate that our method produces a small seed model capable of unfolding to multiple desired depths, with each configuration having performance comparable to that of individually trained models of equal complexity. The contributions of this paper are as follows.

\begin{itemize}
    \item We present a memory-efficient approach for model size compression through a small-to-large perspective, enabling enhancing the performance of a small model by flexibly reusing core structures without increasing physical memory usage.
    \item We introduce an effective training framework that allows models to (un)fold to arbitrary logical depths, addressing the diverse needs of edge devices while remaining mindful of memory constraints during compression.
    \item We conduct experiments on both supervised learning-based Conformer models and self-supervised learning-based speech foundation models, demonstrating the superior compression performance of our methods, compared to state-of-the-art ASR model compression techniques.
\end{itemize}

\vspace{-0.3cm}
\section{Speech Models}
\vspace{-0.1cm}
For supervised learning cases, we consider the convolution-augmented Transformer (Conformer~\cite{gulati2020conformer}), an end-to-end (E2E) ASR model. A Conformer encoder comprises multiple stacking blocks\footnote{We use ``layers" and ``blocks" interchangeably when referring to the repeatedly stacked model structures, e.g., Conformer blocks.}, where each block is composed of the following modules in sequence: a feed-forward module (FFN), a self-attention module (MHSA), a convolution module (Conv), and a second FFN module. An effective way to train an attention-based encoder-decoder (AED) Conformer system is using multitask criterion interpolation~\cite{watanabe2017hybrid} between the loss of CTC and the attention Decoder (i.e., a Transformer decoder). This is given by
%%The overall loss function is:

\begin{equation}
    \setlength\abovedisplayskip{0cm}
    \setlength\belowdisplayskip{0cm}
% \scriptsize
    \mathcal{L}_{conformer}=(1-\lambda)\mathcal{L}_{attention}+\lambda\mathcal{L}_{ctc},
    \label{eq1}
\end{equation}
where $\lambda$ is a constant and empirically set to 0.3
% throughout the experiment s of 
in this paper.

In contrast to conventional supervised learning methods, the emerging paradigm of self-supervised learning typically trains a foundation model from unlabeled examples to obtain general data representations and fine-tune the model on labeled data. Despite the different training paradigms, speech SSL models such as wav2vec 2.0~\cite{baevski2020wav2vec}, HuBERT~\cite{hsu2021hubert} and WavLM~\cite{chen2022wavlm} share similar Transformer backbones with supervised models. For example, wav2vec2.0 consists of a CNN-based feature extractor and a Transformer encoder. %where each encoder layer is composed of a multi-head self-attention (MHSA) and a position-wise feed-forward (FFN) module.
%Speech SSL models such as wav2vec2.0~\cite{baevski2020wav2vec}, HuBERT~\cite{hsu2021hubert}, and WavLM~\cite{chen2022wavlm} share similar Transformer backbones with supervised models. For example, wav2vec2.0 consists of a CNN based feature extractor and a Transformer Encoder, where each Encoder layer contains an MHSA module and an FFN module.
For the ASR task, we fine-tune a pre-trained wav2vec2.0 model with a pure CTC decoder, and its loss is given by $\mathcal{L}_{w2v2}=\mathcal{L}_{ctc}$.

% In the rest of this paper, we perform compression on the encoder of the Conformer and wav2vec2.0 models, which account for a major portion (e.g., over 80\%) of the model size.

\vspace{-0.2cm}
\section{Methods}
\vspace{-0.1cm}
%There are trends that people are designing neural networks with deeper layers and more parameters, as evidence has shown better modeling capabilities of these deeper and larger models. However, the increasing need for memory is not always available on edge devices with limited computational resources. Furthermore, it is also memory costly to perform compression when needing to deal with large models. To address this, In this section, we present methods that generate small models while also reducing the memory usage for training.

\subsection{Network Structure Unfolding}
\vspace{-0.1cm}
% The advantage of small models lies in their low memory consumption, which is user-friendly for the edge device, while they may have inferior performance compared to larger and deeper models. To benefit from both the memory efficiency of small models and the better modeling capacity of deeper models, we present a network structure unfolding approach that allows a small model to reuse its network structures to enhance the model capability and maintain a low memory consumption. 

To benefit from both the memory efficiency of small models and the better modeling capacity of deeper models, we present a network structure unfolding approach that allows a small model to reuse its network structures to enhance the model capability and maintain a low memory consumption. 

\begin{figure}[t]
    \centering
    \includegraphics[width=0.8\linewidth]{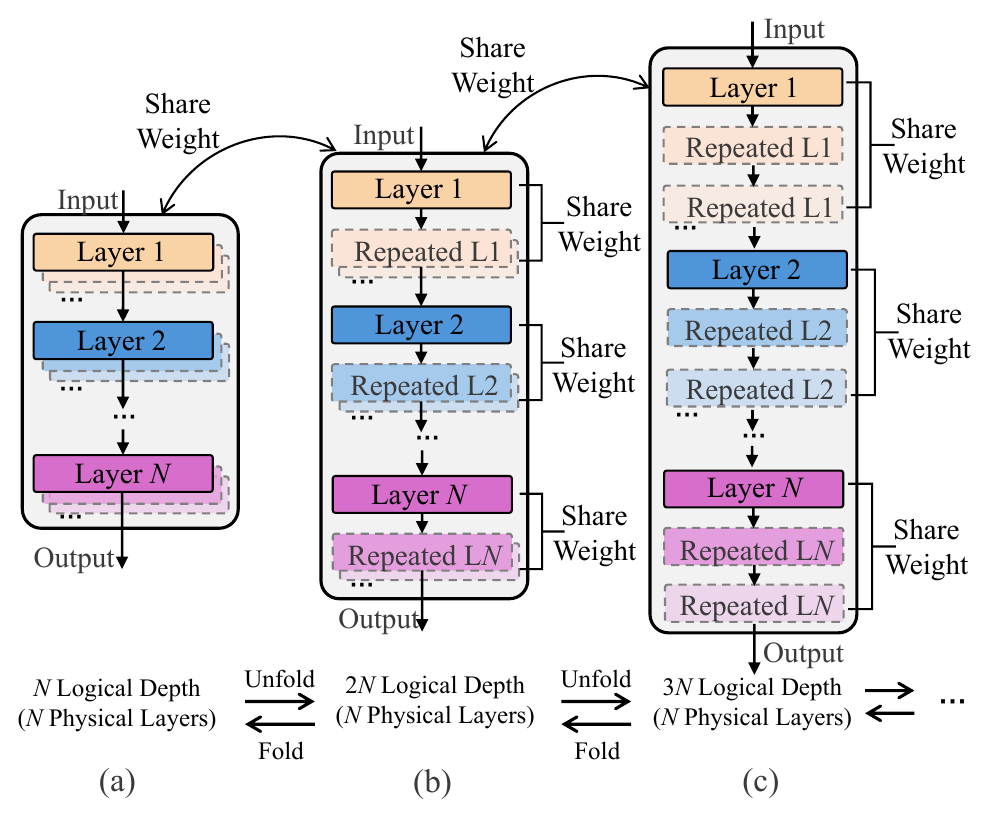}
    \vspace{-0.4cm}
    \caption{Diagram of a foldable network. Systems (a), (b), and (c) share the same $N$ physical layers, and they can (un)fold to each other by changing the number of repeating times of the physical layers, without expanding parameters and memory.}
    \label{fig1}
    \vspace{-0.8cm}
\end{figure}

Suppose we have a small model with $N$ layers (e.g., repeated network structures such as Conformer or Transformer blocks). Typically, the logical depth of the transformations applied to an input is the same as the number of physical layers contained by the model, namely $N$, as shown in Fig.~\ref{fig1}~(a). To enhance its modeling capacity, we can unfold the depth of the logical transformations from $N$ to $2N$ simply by reusing and repeating each of the model layers once, without increasing physical memory, as shown in Fig.~\ref{fig1}~(b). Similarly, one can also unfold the logical depth twice or more to potentially obtain more performance gain while maintaining a low physical memory, as shown in Fig.~\ref{fig1}~(c). In fact, it is not necessary to unfold all layers the same number of times, e.g., unfolding some layers once and the other layers twice is also feasible. For simplicity, in this paper, we follow the rule that a layer is allowed to unfold $k+1$ times only if all the other layers have unfolded $k$ or $k+1$ times, where $k\geq0$.

%Another advantage of the unfolding approach is that 
Training an unfolded network is also straightforward and memory-efficient. During training, only the physical layers of the network are loaded into memory, then when forwarding, each layer is repeated a desired number of times to iteratively transform its input, and the parameters of the physical layers are updated through normal gradient descent methods.

\vspace{-0.cm}

\vspace{-0.2cm}
\subsection{Foldable Network}
\vspace{-0.2cm}
% So far, we have been able to produce a model that can be unfolded and executed at a fixed depth. However, to meet various demands of dynamic environments, we would like to produce a seed model that can flexibly unfold to arbitrary logical depths but not just one fixed depth. 

% To achieve this, we propose a framework that allows joint training of models that share physical layers while being executed at different logical depths. Formally, we denote $P_{n_p}$ as a seed model purely containing $n_p$ ($n_p\in\{1,2,\dots\}$) physical layers and executed without unfolding, and $F_{n_f}^{n_p}$ as a (un)foldable system unfolding from the $n_p$ physical layers to $n_f$ logical depth with $n_p<n_f$, where $P_{n_p}$ and $F_{n_f}^{n_p}$ contain and share the same $n_p$ physical layers. Note $n_f$ is the total effective logical depth, but not the number of unfolding times. Then we denote the losses (i.e. described in Section 2) of these two systems as $\mathcal{L}_{P}^{{n_p}}$ and $\mathcal{L}_{F}^{{n_f}}$, respectively. The criterion of our joint training framework can be given by

Previous studies using weight-sharing blocks~\cite{li2019improving,komatsu2022non}, which are based on a manually defined sharing pattern with a \textit{fixed model depth}. Our work aims to learn a compact seed model that can be flexibly unfolded to \textit{arbitrary logical depths} corresponding to different model complexities or modeling capacities, but not just one fixed depth, namely a foldable network.

To obtain a foldable model, we propose a framework that allows the joint training of multiple unfolding paths. Formally, we denote $P_{n_p}$ as a seed model purely containing $n_p\in\mathbb{N}$ physical layers and executed without unfolding, and $F_{n_f}^{n_p}$ as a (un)foldable system unfolding from the $n_p$ physical layers to $n_f \in\mathbb{N}$ logical depth with $n_p<n_f$, where $P_{n_p}$ and $F_{n_f}^{n_p}$ share the same $n_p$ physical layers. Note $n_f$ is the total effective logical depth, but not the number of unfolding times. Then we denote the losses (i.e. described in Section 2) of these two systems as $\mathcal{L}_{P}^{{n_p}}$ and $\mathcal{L}_{F}^{{n_f}}$, respectively. The criterion of our joint training framework can be given by
\begin{equation}
\setlength\abovedisplayskip{0.1cm}
\setlength\belowdisplayskip{0.1cm}
    \mathcal{L} = \mathcal{L}_{F}^{N_f} + \alpha_p\mathcal{L}_{P}^{n_p},
    \label{eq2}
\end{equation}
where $N_f$ denotes the maximum logical depth we would like the seed model to unfold to and $\alpha_p$ is a constant scalar. This ensures the model $P_{n_p}$ can be unfolded to $F^{n_p}_{N_f}$, and the model $F^{n_p}_{N_f}$ can also be folded back to $P_{n_p}$. Particularly, we found in our experiments that Eq.~(\ref{eq2}) containing only the two explicit loss terms is enough to also enable all middle systems $F^{n_p}_{n_f}$ with depths between $N_f$ and $n_p$ (i.e., $N_f<n_f<n_p$), which indicates the compact seed model with $n_p$ physical layers can unfold to various logical depths to meet different on-device demands. This significantly facilitates memory efficiency because we neither need training of a huge super-network nor separately compressing and storing models of different depths.

To improve the performance of unfolded models, we %adopt stochastic depth~\cite{huang2016deep} (or layer dropout) and 
incorporate a self-distillation process into the compression cycle. Specifically, during joint training, we use the deepest system $\mathcal{L}_{F}^{{N_f}}$ to guide the smallest (all-physical) seed system $P_{n_p}$, which is achieved by adding an additional regularization term:
\begin{equation}
\setlength\abovedisplayskip{0.1cm}
\setlength\belowdisplayskip{0.1cm}
    \mathcal{L}_{Reg} = D_{KL}(SG(p_{N_f})||p_{n_p}),
    \label{eq3}
\end{equation}
where $D_{KL}$ denotes the Kullback-Leibler divergence, $SG(\cdot)$ denotes the stop-gradient operation, and $p_{N_f}$ and $p_{n_p}$ denote the output distribution of system $F^{n_p}_{N_f}$ and $P_{n_p}$, respectively. Then, the final criterion becomes:
\begin{equation}
\setlength\abovedisplayskip{0.1cm}
\setlength\belowdisplayskip{0.1cm}
    \mathcal{L} = \mathcal{L}_{F}^{N_f} + \alpha_p\mathcal{L}_{P}^{n_p} + \alpha_{kl}\mathcal{L}_{Reg},
    \label{eq4}
\end{equation}
where $\alpha_{kl}$ is a constant scalar. It should be noted that the regularization term is computed concurrently within the joint training process, incurring negligible memory and time costs.
%\subsection{Enhancing }

\vspace{-0.2cm}
\section{Experiments}
\vspace{-0.1cm}
\subsection{Experimental Setup}
\vspace{-0.1cm}
\noindent\textbf{Models and Data.}
For supervised learning, we take Conformer AED system configured with the ESPnet~\cite{watanabe2018espnet} recipe\footnote{\href{https://github.com/espnet/espnet/tree/master/egs/swbd/asr1}{ESPnet: egs/swbd/asr1/run.sh}}. We train Conformer with 300-hr Switchboard corpus~\cite{godfrey1992switchboard} and evaluate on NIST Hub5’00, RT02, and RT03 evaluation sets. For SSL cases, the wav2vec2-base-100h/HuBERT-base are downloaded from Huggingface\footnote{facebook/wav2vec2-base-100h; facebook/hubert-base-ls960}. We finetune SSL models on the LibriSpeech 100-hour clean subset~\cite{panayotov2015librispeech}. The baselines are also fine-tuned with 20 epochs for comparison. All baseline models contain a 12-layer encoder.%we compress \red{wav2vec2-base} pre-trained model on HuggingFace Transformer.

\noindent\textbf{Training.}
The Conformer systems are trained from scratch with the same training schedule as the baseline Conformer recipe on a single NVIDIA A40 (48G) GPU. We fine-tune SSL systems with a batch size of 8 for 20 epochs on a single NVIDIA V100 (32G) GPU, where the AdamW optimizer is used with a learning rate of 5e-5 with a linear decay to zero.

\noindent\textbf{Hyper parameters.}
For Conformer, the layer dropout rate of 12 encoder layers increases linearly to 0.1. For SSL systems, the layer dropout rate is set to 0.1. All baseline systems apply the same settings. For joint training, we set $\alpha_p=0.25$ and $\alpha_{kl}=0.1$ for Conformer, and $\alpha_p=0.7$ and $\alpha_{kl}=0.005$ for SSL systems to balance the different terms in the losses base on the observations of their magnitudes, respectively.

\vspace{-0.2cm}
\subsection{Foldable Conformer}
\begin{figure}[ht]
    \centering
    \includegraphics[width=0.8\linewidth]{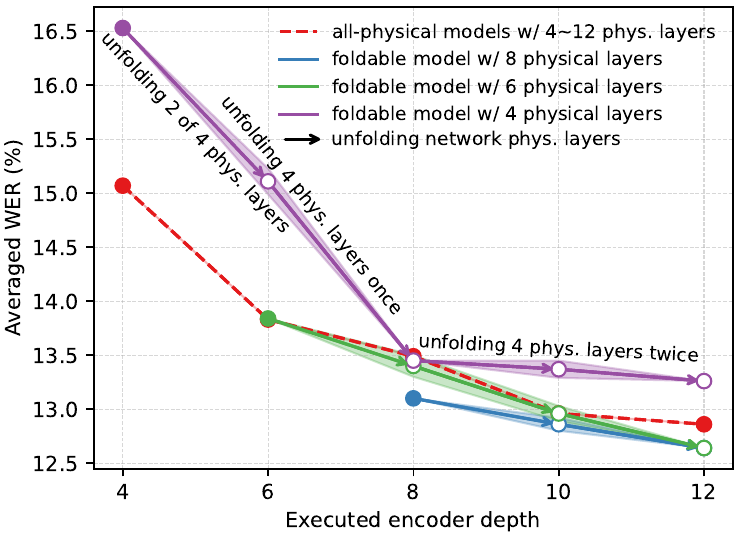}
    \vspace{-0.3cm}
    \caption{Averaged WER of Conformer systems versus executed encoder depths (\# physical layers + unfolded depths). The foldable and all-physical models are the same ones in Table~\ref{tab:1}. Solid circles denote models that purely contain physical layers, while hollow circles denote foldable models unfolded from corresponding all-physical systems. Colored areas cover the ranges of standard deviation across different unfolding paths.}
    \label{fig2}
    \vspace{-0.5cm}
\end{figure}

\vspace{-0.1cm}
We first train Conformer systems with a foldable encoder containing 8, 6, and 4 physical layers, respectively, each of which can be unfolded to arbitrary logical depths with a maximum complexity of 12 layers (i.e., the number of physical layers contained in a baseline system). For the 8-layer foldable Conformer, we only allow it to unfold the last 4 layers since it is enough to reach the logical depth of 12. We compare the performance of the three foldable Conformer models executed at different logical depths with that of individually trained all-physical Conformer models of equal depth. The results are plotted in Fig.~\ref{fig2}, where several trends can be found: \textbf{1)} Applying our network unfolding approach, a small seed model can be enhanced, where the deeper the model unfolds, the lower the WER it produces, which is consistently demonstrated by the 8-, 6-, and 4-layer foldable Conformer systems. \textbf{2)} For the 8-layer and 6-layer foldable Conformer models, the performance of them unfolding to different logical depths (e.g., 6, 8, 10, and 12) is comparable to that of individually constructed models of equal depths, while eliminating the need of separate training and storing these systems of different depths, demonstrating the efficacy and memory efficiency of our approach. \textbf{3)} The foldable models have robust performance. In some cases, there may be more than one path to unfold a model to a desired depth. For example, following the unfolding rule mentioned in Section 3.1, to unfold a 6-layer model to the depth of 8, there are in total 15 combinations. However, in our experiments, we found the performance of different unfolding paths did not deviate much from each other, which is demonstrated by the standard deviation range we plot in Fig.~\ref{fig2}. This indicates, with our methods, one can unfold a model by arbitrarily selecting a path and the performance is guaranteed. \textbf{4)} For the 4-layer foldable Conformer, the WER reduction is more obvious when the physical layers are repeated once than when repeated twice, which suggests it is better to first unfold all the physical layers once before unfolding any of them twice.

\vspace{-0.3cm}

\vspace{-0.2cm}
\begin{table}[ht]
    \centering
    \caption{WER($\downarrow$) of foldable Conformer versus individually compressed systems. A ``*" mark in the last column denotes no statistically significant (MAPSSWE~\cite{gillick1989some}, $\alpha$=0.05) WER increase is observed over the 12-layer baseline system (Sys.~1).}
    \vspace{-0.2cm}
    \setlength\tabcolsep{3pt}
    \resizebox{\linewidth}{!}{
    \begin{tabular}{l|c|c|c|ccccccc|c}
        % \hline
        % \hline
        \toprule
        \multirow{2}{*}{Sys.} & \multirow{2}{*}{\shortstack{\# of physical layers \\ loaded in memory}}& \multirow{2}{*}{\shortstack{Inference \\ depth}} & \multirow{2}{*}{\shortstack{\# of param.\\millions}} & \multicolumn{2}{c}{Hub5’00} & \multicolumn{3}{|c|}{RT02} & \multicolumn{2}{c|}{RT03} & {\multirow{2}{*}{Avg.}}\\
        \cline{5-11}
         & & & &swbd & calhm & \multicolumn{1}{|c}{swbd1} & swbd2 & \multicolumn{1}{c|}{swbd3} & swbd & fsh & \\
         % \hline
         \midrule
         \multicolumn{9}{c}{Baseline (gray) and individually constructed all-physical systems of different depths} & \multicolumn{3}{c}{}\\
         % \hline
         \midrule
         \rowcolor{gray!30}
         1 &12&12 & 44.6 & 7.4 &15.2 &8.9&13.0&15.8&10.6&16.7&12.86\\
         2 &10&10 & 39.3 & 7.6 & 15.5 & 8.9&12.7&16.1&10.6&16.8&12.96$^*$\\
         3 &8&8 & 34.0 & 7.7 &15.9 &9.7&13.4&16.5 &11.2&17.5&13.49  \\
         4 &6&6 & 28.8 & 7.7 &16.3 &9.4&13.9&17.0 &11.4&18.3&13.83  \\
         5 &4&4 & 23.5 & 8.8 &17.6 &10.4&14.8&18.5&12.3&19.6 &15.07  \\
         % \hline
         \midrule
         \multicolumn{4}{c}{Our Method (foldable Conformer)} & \multicolumn{5}{c}{}\\
         % \hline
         \midrule
         E1 & \multirow{3}{*}{\shortstack{8\\(shared)}} & 12 (unfolded) & \multirow{3}{*}{\shortstack{34.0\\(shared)}} & 7.2 &14.9 &8.7&13.0&15.7&10.3&16.5 &12.64$^*$\\
         E2 & & 10 (unfolded) & & 7.4 &15.2 &9.1&13.1&15.9&10.5&16.8 &12.86$^*$\\
         E3 & & 8 & & 7.5 &15.1 &9.3&13.5&16.2&10.6&17.1 &13.10  \\
         % \hline
         \hline
         % \midrule
         S1 & \multirow{4}{*}{\shortstack{6\\(shared)}} & 12 (unfolded) & \multirow{4}{*}{\shortstack{28.8\\(shared)}} & 7.3 &14.9 &8.7&12.6&15.1&10.5&16.8 &12.64$^*$\\
         S2 & & 10 (unfolded) & & 7.6 &15.2 &9.2&13.2&15.9&10.5&16.8 &12.96$^*$\\
         S3 & & 8 (unfolded) & & 7.8 &15.5 &9.2&13.5&16.3&11.0&17.7 &13.40  \\
         S4 & & 6 & & 8.0 &16.5 &9.5&14.1&17.0&11.4&17.9 &13.84  \\
         % \hline
         \hline
         % \midrule

         F1 & \multirow{5}{*}{\shortstack{4\\(shared)}} & 12 (unfolded) & \multirow{5}{*}{\shortstack{23.5\\(shared)}} & 7.5 &15.7 &9.1&13.1&15.7&11.2&17.7 &13.26 \\
         F2 & & 10 (unfolded) & & 7.7 &15.6 &9.3&13.3&16.4&11.1&17.5 &13.37  \\
         F3 & & 8 (unfolded) & & 7.7 &15.7 &9.3&13.2&16.5&11.1&17.6 &13.45  \\
         F4 & & 6 (unfolded) & & 8.6 &18.2 &10.2&15.0&18.8&12.6&19.6 &15.11  \\
         F5 & & 4 & & 9.4 &19.5 &11.7&16.6&20.4&13.9&21.3 &16.53  \\
         % \hline
         % \hline
         \bottomrule

    \end{tabular}
        }
    
    \label{tab:1}
    \vspace{-0.2cm}
\end{table}

The results\footnote{For simplicity, we mainly reported results of models with even numbers of depth while similar trends can also be found in odd cases.} are also reported in Table~\ref{tab:1}. For foldable systems that have multiple possible paths to unfold, we representatively report the system with the median WER. We obtained a 6-layer foldable Conformer that achieved compression with a 35\% model parameter reduction while incurring no statistically significant WER increase when executed at a depth of 12, compared to the 12-layer baseline model (i.e., Sys.~S1 vs. Sys.~1 in Table~\ref{tab:1}). More importantly, the 6-layer foldable Conformer can unfold to various depths and provide comparable performance to those of individually compressed all-physical systems of equal depths while requiring only minimal memory to load the 6 physical layers once for both training and inference. (i.e., Sys.~S1$\sim$S4 vs. Sys.~1$\sim$4).

\vspace{-0.cm}
\begin{figure}[ht]
    \centering
    \includegraphics[width=0.8\linewidth]{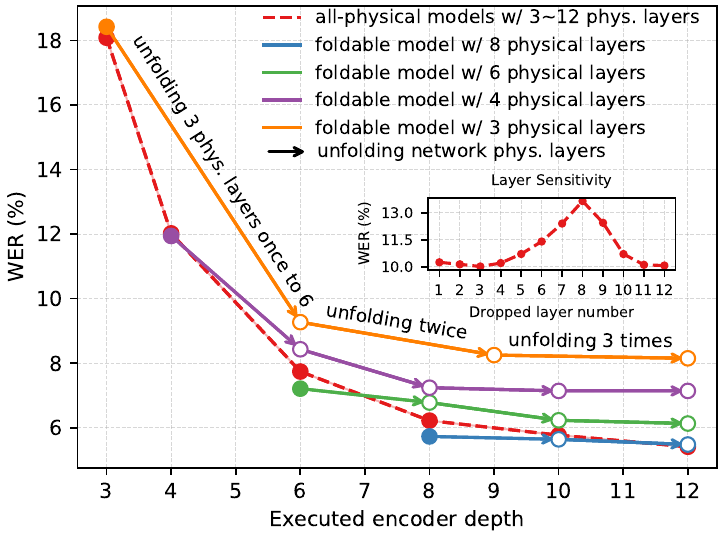}
    \vspace{-0.2cm}
    \caption{WER (test-clean) of wav2vec2 versus executed encoder depths. The foldable and compact models are the same ones in Table~\ref{tab:2}. Solid circles denote systems purely containing physical layers, while hollow circles denote foldable systems unfolded from corresponding all-physical systems.}
    \label{fig3}
    \vspace{-0.3cm}
\end{figure}

\begin{table}[ht]
    \centering
    \caption{WER($\downarrow$) of foldable wav2vec2 systems versus individually compressed systems. A ``*" mark in the last column denotes no statistically significant (MAPSSWE~\cite{gillick1989some}, $\alpha$=0.05) WER increase is observed over the 12-layer baseline system (Sys.~1).}
    \vspace{-0.2cm}
    \setlength\tabcolsep{5pt}
    \resizebox{\linewidth}{!}{
    \begin{tabular}{l|c|c|c|cccc}
        % \hline
        % \hline
        \toprule
        \multirow{2}{*}{Sys.} & \multirow{2}{*}{\shortstack{\# of physical layers\\loaded in memory}}& \multirow{2}{*}{\shortstack{Inference \\ depth}} & \multirow{2}{*}{\shortstack{\# of parameters\\millions}} & \multicolumn{2}{c}{test} & \multicolumn{2}{|c}{dev} \\
        \cline{5-8}
         & & & &clean & other & \multicolumn{1}{|c}{clean} & other \\
         % \hline
         \midrule
         \multicolumn{6}{c}{Baseline (gray) and individually constructed all-physical systems of different depths} & \multicolumn{2}{c}{}\\
         % \hline
         \midrule
         \rowcolor{gray!30}
         1 &12 &12 & 94.4 & 5.41 & 13.88 & 5.38&14.20\\
         2 &10 &10 & 80.2 & 5.77 & 14.55 & 5.60&14.96  \\
         3 &8 &8 & 66.0 & 6.22 &16.91 &6.02&16.55  \\
         4 &6 &6 & 51.9 & 7.74 &23.49 &7.44&22.29  \\
         5 &4 &4 & 37.7 & 12.02 &34.46 &11.41&32.82  \\
         6 &3 &3 & 30.6 & 18.08 &45.55 &17.78&42.58  \\
         % \hline
         \midrule
         \multicolumn{3}{c}{Our Method (foldable wav2vec2)} & \multicolumn{5}{c}{}\\
         % \hline
         \midrule
         E1 & \multirow{3}{*}{\shortstack{8\\(shared)}} & 12 (unfolded) & \multirow{3}{*}{\shortstack{66.0\\(shared)}} & 5.48$^*$ &14.72 &5.41$^*$&14.34$^*$\\
         E2 & & 10 (unfolded) & & 5.64 &15.33 &5.50&15.02  \\
         E3 & & 8 & & 5.73 &15.97 &5.52&15.85  \\
         % \hline
         \hline
         % \midrule
         S1 & \multirow{4}{*}{\shortstack{6\\(shared)}} & 12 (unfolded) & \multirow{4}{*}{\shortstack{51.9\\(shared)}} & 6.20 &17.52 &5.93&16.82 \\
         S2 & & 10 (unfolded) & & 6.23 &18.52 &6.13&17.67  \\
         S3 & & 8 (unfolded) & & 6.78 &20.50 &6.51&19.88  \\
         S4 & & 6 & & 7.21 &21.66 &6.97&21.96 \\
         % \hline
         \hline
         % \midrule

         F1 & \multirow{5}{*}{\shortstack{4\\(shared)}} & 12 (unfolded) & \multirow{5}{*}{\shortstack{37.7\\(shared)}} & 7.14 &22.10 &6.96&20.86 \\
         F2 & & 10 (unfolded) & & 7.14 &22.45 &6.96&21.33  \\
         F3 & & 8 (unfolded) & & 7.24 &23.34 &7.05&22.20 \\
         F4 & & 6 (unfolded) & & 8.43 &26.61 &8.54&25.52  \\
         F5 & & 4 & & 11.94 &35.08 &11.47&33.28  \\
         % \hline
         \hline
         % \midrule
         T1 & \multirow{4}{*}{\shortstack{3\\(shared)}} & 12 (unfolded) & \multirow{5}{*}{\shortstack{30.6\\(shared)}} & 8.15 &26.27 &7.85&24.91 \\
         T2 & & 9 (unfolded) & & 8.25 &26.85 &8.89&25.53  \\
         T3 & & 6 (unfolded) & & 9.27 &30.13 &8.99&28.71 \\
         T4 & & 3 & & 18.42 &46.18 &18.12&43.74  \\
         % \hline
         % \hline
         \bottomrule

    \end{tabular}
        }
    
    \label{tab:2}
    \vspace{-0.6cm}
\end{table}

\vspace{-0.2cm}
\subsection{Foldable SSL Models}
\vspace{-0.1cm}
For SSL pre-trained speech foundation models, we develop a simple way to apply our method by first dropping several physical layers and then unfolding the rest, which ensures minimal memory requirements. To decide which layers to drop, we first evaluate the performance sensitivity of each layer in the wav2vec2 model, which is obtained by respectively testing the model WER when dropping each layer. As shown in the subfigure of Fig.~\ref{fig3}, a larger WER suggests the layer is more sensitive and thus of more importance. Based on this, we follow the rule that a layer corresponding to a lower WER will be dropped with a higher priority. Then, we train foldable wav2vec2 with 8, 6, 4, and 3 physical layers (i.e., the rest layers after dropping), respectively, and plot in Fig.~\ref{fig3}, where similar trends can also be found as those of Conformer models in Fig.~\ref{fig2}, while the standard deviations across different unfolding paths become even too small to show in the figure. Then we show the WERs in Table~\ref{tab:2}, where we obtained an 8-layer foldable wav2vec2 model that achieved compression with a 30\% model parameter reduction while incurring no statistically significant WER increase when executed at a depth of 12, compared to the 12-layer baseline model (i.e., Sys.~E1 vs. Sys.~1 in Table~\ref{tab:2}).

To demonstrate superior compression performance, we further compare our approach with previous compression methods for speech foundation models. We tune foldable wav2vec2/HuBERT with 3 and 2 physical layers (after dropping) and unfold them to 12 layers, denoted as $F_{12}^{3}$ and $F_{12}^{2}$, respectively. As shown in Table~\ref{tab:3}, our method significantly outperforms other methods with an obvious WER reduction under similar model sizes. Remarkably, our method needs memory only to load and train an ultra-small seed model of 2 or 3 physical layers with only a small amount of training data of 100 hours to enable its superior performance and flexible inference depths.

%\subsection{Comparison with Other SSL Compression Methods}

\vspace{-0.3cm}
\begin{table}[ht]
    \centering
    \caption{Foldable SSL systems versus previous SSL model compression methods under comparable model size.}
    \vspace{-0.3cm}
    \setlength\tabcolsep{3pt}
    \resizebox{\linewidth}{!}{
    \begin{tabular}{cccccc}
        % \hline
        %  \hline
        \toprule
         \multirow{2}{*}{Method} & Models need to load & \# supported depths & Hours of& \multirow{2}{*}{\shortstack{\# Parameters\\millions}}&\multirow{2}{*}{\shortstack{WER ($\downarrow$)\\test-clean}}\\
         & in memory when training & when inference &train set& & \\
         % \hline
         % \hline
         \midrule
         DistilHuBERT~\cite{distilhubert} &12-layer teacher + student&1&960&23.49 &13.37 \\
         3-Layer One~\cite{dist2}&12-layer teacher +  student&1&960 &30.58 &12.23 \\
         FitHuBERT~\cite{dist1}&12-layer teacher +  student&1&960 &22.49 &12.09 \\
         FitW2V2~\cite{dist1}&12-layer teacher +  student&1&960&31.63 &11.44 \\
         12-Layer Half~\cite{dist2}&12-layer teacher +  student&1&960 &26.87 &10.96 \\
         DPHuBERT~\cite{dphubert}&12-layer teacher +  student&1&960 &23.59 &10.47 \\
         Wang et al.~\cite{wang2023task}&12-layer baseline&1&960 &26.57 &10.29 \\
         % \hline
         % \hline
         \midrule
         $\boldsymbol{F_{12}^{2}}$-W2V2 (Ours)& 2-layer compressed model&11 (2$\sim$12)&\textbf{100} &23.52 &9.81 \\
         $\boldsymbol{F_{12}^{2}}$-HuBERT (Ours)& 2-layer compressed model&11 (2$\sim$12)&\textbf{100} &23.52 &9.66 \\
         $\boldsymbol{F_{12}^{3}}$-W2V2 (Ours)& 3-layer compressed model&10 (3$\sim$12)&\textbf{100} &30.61 &\textbf{8.15} \\
         $\boldsymbol{F_{12}^{3}}$-HuBERT (Ours)& 3-layer compressed model&10 (3$\sim$12)&\textbf{100} &30.61 &\textbf{7.92} \\
         % \hline
         % \hline
         \bottomrule
    \end{tabular}
    }
    
    \label{tab:3}
    \vspace{-0.5cm}
\end{table}

\begin{table}[ht]
    \centering
    \caption{Effects of joint training and self-distillation for foldable Conformer. WER are averaged over NIST Hub5’00, RT02, and RT03 evaluation sets. Best results are highlighted in bold.}
    \vspace{-0.3cm}
    \setlength\tabcolsep{5pt}
    \resizebox{\linewidth}{!}{
    \begin{tabular}{l|c|c|l|l|l}
         % \hline
         % \hline
         \toprule
         \multirow{2}{*}{Sys.} & \multirow{2}{*}{\shortstack{\# Phys.\\layers}}& \multirow{2}{*}{\shortstack{\# Param.\\millions}}& \multirow{2}{*}{\shortstack{Training loss}} & \multirow{2}{*}{\shortstack{Inference depth}}&\multirow{2}{*}{\shortstack{Avg. WER ($\downarrow$)\\(relative increase)}}\\
         & & & & \\
         % \hline
         % \hline
         \midrule
         0 & \multirow{2}{*}{12} & \multirow{2}{*}{44.6} & \multirow{2}{*}{$\mathcal{L}_{P}^{12}$}& {\cellcolor{gray!30}12 -- Baseline} & {\cellcolor{gray!30}12.86 (+0.0 \%)}\\
         1 & & & & 8 -- Dropping last 4 layers & 21.50 (+67.2 \%)\\
         \hline
         % \midrule
         2 & \multirow{2}{*}{8} & \multirow{2}{*}{34.0} & \multirow{2}{*}{$\mathcal{L}_{P}^{8}$}& 12 -- Unfolding last 4 layers & 17.50 (+36.1 \%)\\
         3 & & & & 8 & 13.45 (+4.6 \%)\\
         \hline
         % \midrule
         4 & \multirow{2}{*}{\shortstack{8}}&\multirow{2}{*}{34.0} & \multirow{2}{*}{$\mathcal{L}_{F}^{12}$}  & 12 -- Unfolding last 4 layers & 12.88 (+0.2 \%)\\
         5 & & & &8 & 27.23 (+111.7 \%)\\
         % \hline
         % \hline
         % \midrule
         \midrule
         E0 & \multirow{2}{*}{\shortstack{8}} & \multirow{2}{*}{34.0}&\multirow{2}{*}{$\mathcal{L}_{F}^{12}+\mathcal{L}_{P}^{8}$} & 12 -- Unfolding last 4 layers & 12.90 (+0.3 \%)\\
         E1 & & & & 8 & 13.71 (+6.6 \%)\\
         \hline
         % \midrule
         K0 & \multirow{2}{*}{\shortstack{8}} & \multirow{2}{*}{34.0} &\multirow{2}{*}{$\mathcal{L}_{F}^{12}+\mathcal{L}_{P}^{8}+\mathcal{L}_{Reg}$}& 12 -- Unfolding last 4 layers & \textbf{12.64 (-1.7 \%)}\\
         K1 & & & & 8 & \textbf{13.10 (+1.9 \%)}\\
         % \hline
         % \hline
         \bottomrule
    \end{tabular}
    }
    \label{tab:4}
    \vspace{-0.6cm}
\end{table}

\vspace{-0.1cm}
\subsection{Ablation Study}
\vspace{-0.2cm}
This section studies the effectiveness of our approach (i.e., the joint training and KL regularization). We first show that individually constructed systems cannot be (un)folded. As shown in Table~\ref{tab:4} from Sys.~0 to Sys.~5, one can see that systems trained with a single loss can only be executed at the same depth as in training, while incurring significant WER increases when executed at other depths. For example, Sys.~0 and 1 contain 12 layers and are trained at a depth of 12. However, when inference, it works only normally at the depth of 12, while resulting in a relative WER increase of 67.2\% if folded to the depth of 8. This is also the case for the 8-layer system that trained at the depth of 8 but can not be unfolded (i.e., Sys.~2 and 3), and the 8-layer system that trained at the unfolded depth of 12 that can not be folded back (i.e., Sys.~4 and 5). In contrast, by using our joint training framework, the 8-layer foldable system (i.e., Sys.~E0 and E1) can be executed at depths of both 8 and 12. Actually, it can be executed at arbitrary depths between 8 and 12 as we already showed in the previous sections. Finally, when the KL-regularization term is added, the performance of the foldable model is further improved at any depth (i.e., Sys.~K0 and K1 vs. Sys.~E0 and E1, respectively).

\vspace{-0.3cm}
\section{Conclusion}
\vspace{-0.1cm}

% This paper presents a novel architecture compression approach for ASR systems from a small-to-large perspective to effectively address key challenges of memory efficiency, performance, and scalability. Our model's capability to unfold to arbitrary depths ensures adaptability to various edge device requirements while reducing memory and time consumption to develop and maintain multiple larger systems. Through our training framework and self-distillation process, we achieve performance comparable to individually trained models across multiple configurations. Experimental results validate the efficacy of our approach, contributing to the advancement of resource-efficient ASR systems in constrained environments.

This paper introduces a novel architecture compression approach for ASR systems to effectively address essential challenges related to memory efficiency, model performance, and adaptability to various edge device requirements. It enables the development of a single compact model that can unfold to various logical depths, achieving performance comparable to that of individually constructed models while demonstrating superior results compared to previous ASR model compression methods. The model’s capability to unfold to arbitrary depths enhances its versatility while minimizing memory usage and development time associated with managing multiple larger systems, contributing to the development of resource-efficient ASR systems suitable for constrained environments.

\newpage

\section{Acknowledgements}
This research is supported by Hong Kong RGC GRF grant No. 14200220, 14200021, 14200324, Innovation Technology Fund grant No. ITS/218/21, and Youth Innovation Promotion Association CAS Grant (2023119).

\bibliographystyle{IEEEtran}
\bibliography{mybib}

% Generated by IEEEtran.bst, version: 1.13 (2008/09/30)
\begin{thebibliography}{10}
\providecommand{\url}[1]{#1}
\csname url@samestyle\endcsname
\providecommand{\newblock}{\relax}
\providecommand{\bibinfo}[2]{#2}
\providecommand{\BIBentrySTDinterwordspacing}{\spaceskip=0pt\relax}
\providecommand{\BIBentryALTinterwordstretchfactor}{4}
\providecommand{\BIBentryALTinterwordspacing}{\spaceskip=\fontdimen2\font plus
\BIBentryALTinterwordstretchfactor\fontdimen3\font minus \fontdimen4\font\relax}
\providecommand{\BIBforeignlanguage}[2]{{%
\expandafter\ifx\csname l@#1\endcsname\relax
\typeout{** WARNING: IEEEtran.bst: No hyphenation pattern has been}%
\typeout{** loaded for the language `#1'. Using the pattern for}%
\typeout{** the default language instead.}%
\else
\language=\csname l@#1\endcsname
\fi
#2}}
\providecommand{\BIBdecl}{\relax}
\BIBdecl

\bibitem{baevski2020wav2vec}
A.~Baevski, Y.~Zhou \emph{et~al.}, ``wav2vec 2.0: A framework for self-supervised learning of speech representations,'' in \emph{NeurIPS}, 2020.

\bibitem{hsu2021hubert}
W.-N. Hsu, B.~Bolte \emph{et~al.}, ``{HuBERT}: Self-supervised speech representation learning by masked prediction of hidden units,'' \emph{IEEE/ACM T-ASLP}, vol.~29, pp. 3451--3460, 2021.

\bibitem{baevski2022data2vec}
A.~Baevski, W.-N. Hsu \emph{et~al.}, ``Data2vec: A general framework for self-supervised learning in speech, vision and language,'' in \emph{ICML}, 2022.

\bibitem{chen2022wavlm}
S.~Chen, C.~Wang \emph{et~al.}, ``{WavLM}: Large-scale self-supervised pre-training for full stack speech processing,'' \emph{IEEE J-STSP}, vol.~16, no.~6, pp. 1505--1518, 2022.

\bibitem{quant1}
S.~Ding, P.~Meadowlark \emph{et~al.}, ``4-bit conformer with native quantization aware training for speech recognition,'' in \emph{Interspeech}, 2022.

\bibitem{quant2}
S.~Kim, A.~Gholami \emph{et~al.}, ``Integer-only zero-shot quantization for efficient speech recognition,'' in \emph{ICASSP}, 2022.

\bibitem{quant3}
O.~Rybakov, P.~Meadowlark \emph{et~al.}, ``2-bit conformer quantization for automatic speech recognition,'' in \emph{Interspeech}, 2023.

\bibitem{xu2025effective}
H.~Xu, Z.~Li \emph{et~al.}, ``Effective and efficient mixed precision quantization of speech foundation models,'' in \emph{ICASSP}, 2025.

\bibitem{rathod2022multi}
J.~Rathod, N.~Dawalatabad \emph{et~al.}, ``Multi-stage progressive compression of conformer transducer for on-device speech recognition,'' in \emph{Interspeech}, 2022.

\bibitem{distilhubert}
H.-J. Chang, S.-w. Yang \emph{et~al.}, ``{DistilHuBERT}: Speech representation learning by layer-wise distillation of hidden-unit bert,'' in \emph{ICASSP}, 2022.

\bibitem{dist1}
Y.~Lee, K.~Jang \emph{et~al.}, ``{FitHuBERT}: Going thinner and deeper for knowledge distillation of speech self-supervised learning,'' in \emph{Interspeech}, 2022.

\bibitem{dist2}
T.~Ashihara, T.~Moriya \emph{et~al.}, ``Deep versus wide: An analysis of student architectures for task-agnostic knowledge distillation of self-supervised speech models,'' in \emph{Interspeech}, 2022.

\bibitem{dist3}
Y.~Fu, Y.~Kang \emph{et~al.}, ``{DistillW2V2}: A small and streaming wav2vec 2.0 based asr model,'' \emph{arXiv preprint arXiv:2303.09278}, 2023.

\bibitem{de2023distilling}
D.~de~Oliveira and T.~Gerkmann, ``Distilling {HuBERT} with {LSTMs} via decoupled knowledge distillation,'' in \emph{ICASSP}, 2024.

\bibitem{park2023conformer}
J.~Park, S.~Jin \emph{et~al.}, ``Conformer-based on-device streaming speech recognition with {KD} compression and two-pass architecture,'' in \emph{IEEE SLT}, 2023.

\bibitem{pru3}
Z.~Wu, D.~Zhao \emph{et~al.}, ``Dynamic sparsity neural networks for automatic speech recognition,'' in \emph{ICASSP}, 2021.

\bibitem{pru4}
C.-I.~J. Lai, Y.~Zhang \emph{et~al.}, ``{PARP}: Prune, adjust and re-prune for self-supervised speech recognition,'' in \emph{NeurIPS}, 2021.

\bibitem{pru5}
J.~Lee, J.~Kang \emph{et~al.}, ``Layer pruning on demand with intermediate ctc,'' in \emph{Interspeech}, 2021.

\bibitem{dphubert}
Y.~Peng, Y.~Sudo \emph{et~al.}, ``{DPHuBERT}: Joint distillation and pruning of self-supervised speech models,'' in \emph{Interspeech}, 2023.

\bibitem{jiang2023accurate}
H.~Jiang, L.~L. Zhang \emph{et~al.}, ``Accurate and structured pruning for efficient automatic speech recognition,'' in \emph{Interspeech}, 2023.

\bibitem{lodagala2023pada}
V.~S. Lodagala, S.~Ghosh \emph{et~al.}, ``{PADA}: Pruning assisted domain adaptation for self-supervised speech representations,'' in \emph{IEEE SLT}, 2023.

\bibitem{hj}
Y.~Peng, K.~Kim \emph{et~al.}, ``Structured pruning of self-supervised pre-trained models for speech recognition and understanding,'' in \emph{ICASSP}, 2023.

\bibitem{wang2023task}
H.~Wang, S.~Wang \emph{et~al.}, ``Task-agnostic structured pruning of speech representation models,'' in \emph{Interspeech}, 2023.

\bibitem{gu2024sparsewav}
T.~Gu, B.~Liu \emph{et~al.}, ``{SparseWAV}: Fast and accurate one-shot unstructured pruning for large speech foundation models,'' in \emph{Interspeech}, 2024.

\bibitem{lr1}
D.~Povey, G.~Cheng \emph{et~al.}, ``Semi-orthogonal low-rank matrix factorization for deep neural networks,'' in \emph{Interspeech}, 2018.

\bibitem{lr2}
S.~Hu, X.~Xie \emph{et~al.}, ``Neural architecture search for {LF-MMI} trained time delay neural networks,'' \emph{IEEE/ACM T-ASLP}, vol.~30, pp. 1093--1107, 2022.

\bibitem{lr3}
S.~Li, M.~Xu \emph{et~al.}, ``Efficient conformer-based speech recognition with linear attention,'' in \emph{APSIPA ASC}, 2021.

\bibitem{lilossless}
Z.~Li, T.~Wang \emph{et~al.}, ``Lossless 4-bit quantization of architecture compressed conformer asr systems on the 300-hr switchboard corpus,'' in \emph{Interspeech}, 2023.

\bibitem{lan2019albert}
Z.~Lan, M.~Chen \emph{et~al.}, ``{ALBERT}: A lite bert for self-supervised learning of language representations,'' in \emph{ICLR}, 2020.

\bibitem{gao2021extremely}
Z.~Gao, Y.~Yao \emph{et~al.}, ``Extremely low footprint end-to-end asr system for smart device,'' in \emph{Interspeech}, 2021.

\bibitem{lin2023weight}
G.-T. Lin, Q.~Tang \emph{et~al.}, ``Weight-sharing supernet for searching specialized acoustic event classification networks across device constraints,'' in \emph{ICASSP}, 2023.

\bibitem{hernandez2023sharing}
S.~M. Hernandez, D.~Zhao \emph{et~al.}, ``Sharing low rank conformer weights for tiny always-on ambient speech recognition models,'' in \emph{ICASSP}, 2023.

\bibitem{wang2024residualtransformer}
Y.~Wang and J.~Li, ``Residual{T}ransformer: Residual low-rank learning with weight-sharing for transformer layers,'' in \emph{ICASSP}, 2024.

\bibitem{once}
H.~Cai, C.~Gan \emph{et~al.}, ``Once for all: Train one network and specialize it for efficient deployment,'' in \emph{ICLR}, 2020.

\bibitem{uslim}
J.~Yu and T.~S. Huang, ``Universally slimmable networks and improved training techniques,'' in \emph{ICCV}, 2019.

\bibitem{zhang2021self}
L.~Zhang, C.~Bao \emph{et~al.}, ``Self-distillation: Towards efficient and compact neural networks,'' \emph{IEEE T-PAMI}, vol.~44, no.~8, pp. 4388--4403, 2021.

\bibitem{nagaraja2021collaborative}
V.~Nagaraja, Y.~Shi \emph{et~al.}, ``Collaborative training of acoustic encoders for speech recognition,'' in \emph{Interspeech}, 2021.

\bibitem{wang2022lighthubert}
R.~Wang, Q.~Bai \emph{et~al.}, ``{LightHuBERT}: Lightweight and configurable speech representation learning with once-for-all hidden-unit bert,'' in \emph{Interspeech}, 2022.

\bibitem{li2024one}
Z.~Li, H.~Xu \emph{et~al.}, ``One-pass multiple conformer and foundation speech systems compression and quantization using an all-in-one neural model,'' in \emph{Interspeech}, 2024.

\bibitem{akhtar2023small}
Z.~Akhtar, M.~O. Khursheed \emph{et~al.}, ``Small-footprint slimmable networks for keyword spotting,'' in \emph{ICASSP}, 2023.

\bibitem{gulati2020conformer}
A.~Gulati, J.~Qin \emph{et~al.}, ``Conformer: Convolution-augmented transformer for speech recognition,'' in \emph{Interspeech}, 2020.

\bibitem{watanabe2017hybrid}
S.~Watanabe, T.~Hori \emph{et~al.}, ``Hybrid {CTC/attention} architecture for end-to-end speech recognition,'' \emph{IEEE Journal of Selected Topics in Signal Processing}, vol.~11, no.~8, pp. 1240--1253, 2017.

\bibitem{li2019improving}
S.~Li, R.~Dabre \emph{et~al.}, ``Improving transformer-based speech recognition systems with compressed structure and speech attributes augmentation.'' in \emph{Interspeech}, 2019.

\bibitem{komatsu2022non}
T.~Komatsu, ``Non-autoregressive asr with self-conditioned folded encoders,'' in \emph{ICASSP}, 2022.

\bibitem{watanabe2018espnet}
S.~Watanabe, T.~Hori \emph{et~al.}, ``Espnet: End-to-end speech processing toolkit,'' \emph{arXiv preprint arXiv:1804.00015}, 2018.

\bibitem{godfrey1992switchboard}
J.~J. Godfrey, E.~C. Holliman \emph{et~al.}, ``{SWITCHBOARD}: Telephone speech corpus for research and development,'' in \emph{ICASSP}, 1992.

\bibitem{panayotov2015librispeech}
V.~Panayotov, G.~Chen \emph{et~al.}, ``{LibriSpeech}: an asr corpus based on public domain audio books,'' in \emph{ICASSP}, 2015.

\bibitem{gillick1989some}
L.~Gillick and S.~J. Cox, ``Some statistical issues in the comparison of speech recognition algorithms,'' in \emph{ICASSP}, 1989.

\end{thebibliography}

\end{document}